\pdfoutput=1
\documentclass[a4paper, 11pt, oneside]{article}

\usepackage{epsf,amsmath,amssymb,graphicx,dcolumn,slashed}
\usepackage{scalefnt,cite,hyperref,bbold}
\usepackage{cleveref,etoolbox,color,soul}
\usepackage{listings,booktabs}
\usepackage{epstopdf}
\usepackage{amsfonts}

\newcommand{\half}{\tfrac{1}{2}}

\newlength{\absize}

\newcommand{\beq}{\begin{equation}}  
\newcommand{\eeq}{\end{equation}}
\newcommand{\bea}{\begin{eqnarray}} 
\newcommand{\eea}{\end{eqnarray}}

\newcounter{notecount}
\numberwithin{equation}{section}

\begin{document}

\long\def\symbolfootnote[#1]#2{\begingroup%
\def\thefootnote{\fnsymbol{footnote}}\footnote[#1]{#2}\endgroup}

\title{
\Large\bf\boldmath
GOOFy fermions
\unboldmath
}
 
\author{
  \addtocounter{footnote}{2}
  P.~M.~Ferreira$^{(1,2)}$\thanks{\tt pmmferreira@fc.ul.pt}
\\[0.4cm]
$^{(1)}\!$
  \small Instituto~Superior~de~Engenharia~de~Lisboa~---~ISEL,
  1959-007~Lisboa, Portugal
  \\[2mm]
  $^{(2)}\!$
  \small Centro~de~F\'\i sica~Te\'orica~e~Computacional,
  Faculdade~de~Ci\^encias, Universidade~de~Lisboa
  }
\maketitle

\begin{abstract}
\noindent
A new class of symmetries, named GOOFy symmetries of two Higgs doublet models was recently discovered, the result of an unorthodox
transformation on scalar and gauge fields and spacetime coordinates. It was explicitly shown that it is 
possible to choose Yukawa matrix textures which respect those symmetries up to two-loops. In this work we will 
establish the fermion field transformations for the two Higgs doublet models to be considered in the context of 
the new symmetries established. We identify two phenomenologically viable models with a GOOFy symmetry, including
fermionic sectors.
\end{abstract}

\setcounter{footnote}{0}

\section{Introduction}

The two Higgs doublet model (2HDM)~\cite{Lee:1973iz}  has long been one of the most popular extensions of the Standard Model (SM) 
of particle  physics. The model has a richer scalar spectrum, containing three neutral and one charged 
particles; the possibility of dark matter candidates; new sources of CP violation from spontaneous or explicit symmetry breaking; 
and more complex Yukawa sectors, including flavour changing neutral currents (FCNC) mediated by scalars (for a review, see~\cite{Branco:2011iw}).
The most general 2HDM scalar potential, however, is much more complex than the SM one -- it depends on 11 independent real parameters (seemingly
14, but field redefinitions mean 3 of those are spurious parameters~\cite{Davidson:2005cw}), as opposed to the 2 parameters characterizing the
SM Higgs potential. It is therefore common to impose global (discrete or continuous) symmetries on the 2HDM to reduce its number of parameters
and increase its predictive power, as well as obtaining interesting phenomenology.  Indeed, the existence of those FCNC interactions in the 
Yukawa sector, which are highly constrained by experimental data, motivated the introduction of a $Z_2$ symmetry by Glashow, 
Weinberg~\cite{Glashow:1976nt} and Paschos~\cite{Paschos:1976ay}. Another example would be the Peccei-Quinn $U(1)$ symmetry~\cite{Peccei:1977hh},
introduced to attempt to explain the strong CP problem. In all there are six independent global symmetries of the $SU(2)\times U(1)$ scalar
potential~\cite{Ivanov:2007de}. These symmetries involve unitary transformations on the doublets -- in which these transform as 
$\phi^\prime_i = U_{ij}\Phi_j$, with $i,j = 1, 2$, for a generic $U \in U(2)$ unitary matrix and a sum on $i$ is assumed -- or their complex conjugates
-- $\phi^\prime_i = X_{ij}\Phi_j^*$, for $X\in U(2)$. The former are called {\em Higgs family symmetries}, the latter {\em generalized CP symmetries}.
Crucially, these symmetries preserve the kinetic terms and gauge interactions of the scalar fields. They also force certain relations between the parameters
of the model which are renormalization group (RG) invariant to all orders of perturbation theory. 

Recently~\cite{Ferreira:2023dke}, however, a new class of (suspected) symmetries was identified, dubbed {\em $r_0$-symmetries} or, due to the
initials of the authors of that paper, GOOFy symmetries. In that paper it was shown that certain 
relations among 2HDM scalar parameters were RG invariant to all orders of perturbation theory, but such relations could not be reproduced by any
of the six symmetry classes hitherto identified. It was shown that RG invariance
held to all orders if scalar and gauge interactions were considered.  This strongly suggests that there is an underlying symmetry of the model 
causing this RG invariance, but such a symmetry cannot be generated by unitary transformations involving scalar doublets, or their complex conjugates,
which leave the kinetic terms invariant, including scalar-gauge interactions. 
In~\cite{Ferreira:2023dke} an unorthodox~\footnote{Notice that in this context the authors
of~\cite{Doring:2024kdg} use the term ``unorthodox" to specifically designate model extensions with non-normal subgroups.} 
proposal for the origin of such RG-invariant relations was made: the scalar potential parameters relations could be obtained if
the real components of the doublets were rescaled by imaginary factors of ``$\pm\, i$"; such relations cause a sign change in 
the scalar kinetic terms,
however, which is then compensated by an imaginary scaling of the spacetime coordinates themselves,
$x_\mu\rightarrow i x_\mu$. An imaginary scaling
of the gauge boson fields is then also required so that the full scalar + gauge lagrangian is invariant under 
these strange transformations. The striking
aspect of this bizarre procedure is how consistent it turns out to be: imaginary scaling of scalar fields requires the imaginary scaling of spacetime 
coordinates to keep the scalar kinetic terms invariant; for gauge-scalar interaction terms to be invariant gauge fields must also be scaled by imaginary 
coefficients, and that scaling is shown to then preserve the gauge fields' kinetic terms and self interactions. Further, in~\cite{Ferreira:2025ate} 
this procedure was shown that invariance under an $r_0$-symmetry could be extended to the 1-loop effective potential.  
The strangeness of this imaginary scaling then justified the dubbing of these symmetries as GOOFy. 

This 2HDM discovery has led to several attempts to explain it without recourse to the imaginary scaling proposed 
in~\cite{Ferreira:2023dke,Ferreira:2025ate}. In~\cite{Haber:2025cbb} a toy model with two real scalars was studied~\cite{Boh:2024MPI}, and shown 
to possess parameter relations which are all-order RG-invariant under a $r_0$-like imaginary scaling of the fields; this inspired a complexification
of the model, whereupon real scalar fields were replaced by complex ones and the larger theory was required to be invariant under ``normal" symmetries. 
The parameter relations protected by symmetries in the complexified theory could then imply the desired relations among the parameters of 
the original model. This proposal implies that symmetry-like relations in a model could be caused by genuine symmetries in a larger, not 
necessarily real, model -- the idea is not that the complexified theory is a UV completion of the original model, rather that the RG-invariant
relations are obtained as algebraic identities shared between both theories. 
In~\cite{Trautner:2025yxz} a different argument was proposed, analysing transformations of doublet fields which do not preserve
the form of the kinetic terms, rather changing their sign; it was then argued that such sign changes 
do not affect the RG-invariant relations found. In this proposal it is argued that GOOFy-violating kinetic terms, including the gauge interaction 
terms present in the covariant derivatives, act as \textit{soft breaking terms} for the RG invariant relations found, and therefore
should not spoil them at any order.

Another aspect to consider is the role of fermions in these RG-invariant relations: in~\cite{Ferreira:2023dke} an all-orders demonstration of 
RG invariance when fermions were included was not found, but for two specific Yukawa matrix textures it was possible to show by explicit
calculation of $\beta$-functions that the $r_0$-symmetry parameter relations were preserved by renormalization up to at least two loops. Those
Yukawa textures were those found for the 2HDM with generalized CP symmetries, namely the CP2 and CP3 models~\cite{Ferreira:2010bm}.
Recently~\cite{Trautner:2025prm,deBoer:2025jhc} the generalization of GOOFy symmetries to fermions was undertaken,
applied to the Standard Model in an attempt to address the hierarchy problem. The extension of GOOFy symmetries 
to fermions was also undertaken in~\cite{Grzadkowski:2026gkx}.

In this work we will take the imaginary scaling of~\cite{Ferreira:2023dke,Ferreira:2025ate} seriously and find its extension to the Yukawa sector.
At the very least, this strange procedure has served as an inspiration for alternative explanations for the RG-invariant relations found, so
maybe its generalization to fermions will also be useful in the same manner. The most exciting possibility, of course, is that these imaginary
field and spacetime coordinate transformations are indeed a new type of symmetry. We will not be able to answer that question in this paper,
unfortunately. We will review the results from~\cite{Ferreira:2023dke} for the scalar + gauge 2HDM GOOFy symmetries and proposed 
imaginary transformations in section~\ref{sec:sg}. 
In section~\ref{sec:yuk} we will propose a GOOFy transformation for the fermions similar to generalized CP transformations, 
and show that under them the
full lagrangian remains invariant. In particular we will show the coherence of the procedure, demonstrating how the new transformations 
leave invariant the fermions' kinetic terms; their gauge interactions; and their interactions with scalars as well. We 
will show how
the GOOFy models of~\cite{Ferreira:2023dke} naturally require that the Yukawa matrices involved have the textures corresponding 
to the CP2 or CP3 models. We will also consider a new GOOFy form of Higgs family-like transformations also generalizable to 
fermions. Two versions of the 2HDM with GOOFy symmetries including the fermion sector are proposed, which are phenomenologically viable
and have simple Yukawa matrix textures.

\section{The Two Higgs Doublet Model and GOOFy symmetries}
\label{sec:sg}

The 2HDM expands the scalar sector of the SM, in that it has two hypercharge $Y = 1$ scalar doublets, $\Phi_1$ and $\Phi_2$, 
instead of just
one. The most general scalar potential invariant under the electroweak gauge group $SU(2)_L\times U(1)_Y$ is written as
\bea
V &=& m_{11}^2\Phi_1^\dagger\Phi_1+m_{22}^2\Phi_2^\dagger\Phi_2
-[m_{12}^2\Phi_1^\dagger\Phi_2+{\rm h.c.}]+\half\lambda_1(\Phi_1^\dagger\Phi_1)^2
+\half\lambda_2(\Phi_2^\dagger\Phi_2)^2
+\lambda_3(\Phi_1^\dagger\Phi_1)(\Phi_2^\dagger\Phi_2)\nonumber\\[8pt]
&&\quad
+\lambda_4(\Phi_1^\dagger\Phi_2)(\Phi_2^\dagger\Phi_1)
+\left\{\half\lambda_5(\Phi_1^\dagger\Phi_2)^2
+\big[\lambda_6(\Phi_1^\dagger\Phi_1)
+\lambda_7(\Phi_2^\dagger\Phi_2)\big]
\Phi_1^\dagger\Phi_2+{\rm h.c.}\right\}\,,
 \label{eq:pot}
\eea
where all parameters are real, except for $m^2_{12}$ and $\lambda_{5,6,7}$. Analyses of possible symmetries
vacuum structure and others are facilitated by rewriting the above potential in terms of the four gauge-invariant
bilinear quantities constructed with the 
doublets~\cite{Velhinho:1994np,Nagel:2004sw,Ivanov:2005hg,Ivanov:2006yq,Ivanov:2007de,Maniatis:2006fs,Maniatis:2006jd,
Nishi:2006tg,Nishi:2007nh,Nishi:2007dv,Maniatis:2007vn}, 
\bea
\begin{array}{rcl}
r_0 &=&
\frac{1}{2}
\left( \Phi_1^\dagger \Phi_1 + \Phi_2^\dagger \Phi_2 \right),
\\*[2mm]
r_1 &=&
\frac{1}{2}
\left( \Phi_1^\dagger \Phi_2 + \Phi_2^\dagger \Phi_1 \right)
= \mbox{Re}\left( \Phi_1^\dagger \Phi_2 \right),
\\*[2mm]
r_2 &=&
- \frac{i}{2}
\left( \Phi_1^\dagger \Phi_2 - \Phi_2^\dagger \Phi_1 \right)
= \mbox{Im} \left( \Phi_1^\dagger \Phi_2 \right),
\\*[2mm]
r_3 &=&
\frac{1}{2}
\left( \Phi_1^\dagger \Phi_1 - \Phi_2^\dagger \Phi_2 \right).
\end{array}
\label{eq:rs}
\eea
The scalar potential may therefore be written as
\beq
V \,=\,M_\mu\,r^\mu\,+\,\Lambda_{\mu\nu}\,r^\mu\,r^\nu\,,
\eeq
with a Minkowski-like formalism requiring the 4-vectors
\bea
r^\mu &=& (r_0\,,\,r_1\,,\,r_2\,,\,r_3) \,=\, (r_0\,,\,\vec{r})\,, \nonumber \\
M^\mu &=& \left(m^2_{11} + m^2_{22}\,,\, 2\mbox{Re}(m^2_{12})
\,,\, -2\mbox{Im}(m^2_{12})\,,\,m^2_{22} - m^2_{11}\right)\,=\, (M_0\,,\,\vec{M})\,,
\label{eq:defbil}
\eea
and the tensor
\begin{align}
	\Lambda^{\mu\nu} & = \begin{pmatrix} \Lambda_{00} & \vec{\Lambda} \\
\vec{\Lambda}^T & \Lambda \end{pmatrix}
	=
	\begin{pmatrix}
		\frac{1}{2}(\lambda_1 + \lambda_2) + \lambda_3 &
	 -\mbox{Re}\left(\lambda_6 + \lambda_7\right) &
	\mbox{Im}\left(\lambda_6 + \lambda_7\right) &
		\frac{1}{2}(\lambda_2 - \lambda_1) \\
		-\mbox{Re}\left(\lambda_6 + \lambda_7\right) &
		\lambda_4 + \mbox{Re} \left( \lambda_5\right) &
		- \mbox{Im} \left( \lambda_5 \right) &
		\mbox{Re}\left(\lambda_6 - \lambda_7\right)
	\\
	\mbox{Im} \left( \lambda_6 + \lambda_7\right)&
	- \mbox{Im} \left( \lambda_5 \right) &
	\lambda_4 - \mbox{Re} \left( \lambda_5\right) &
	- \mbox{Im} \left( \lambda_6 - \lambda_7\right) \\
	\frac{1}{2}(\lambda_2 - \lambda_1) &
	\mbox{Re}\left( \lambda_6 - \lambda_7\right) &
	- \mbox{Im} \left( \lambda_6 - \lambda_7 \right) &
	\frac{1}{2}(\lambda_1 + \lambda_2) - \lambda_3
	\end{pmatrix}\,.
\label{eq:Lambda}
\end{align}
It has been shown~\cite{Ivanov:2007de} that this potential has six independent global symmetries, obtained in this formalism 
via rotations and axis inversions on the vector $\vec{r}$~\footnote{Which correspond, respectively, to unitary transformations mixing either
the doublets, $\Phi_i^\prime = U_{ij} \Phi_j$, or their complex conjugates, $\Phi_i^\prime = X_{ij} \Phi_j^*$, for $U(2)$ matrices
$U$ and $X$.}. In~\cite{Ferreira:2023dke} it was shown, via direct analysis of the model's $\beta$-functions, that the set of relations
\beq
m^2_{11} + m^2_{22}\,=\,0\;\;\; , \;\;\; \lambda_1 = \lambda_2 \;\;,\;\; \lambda_6 = -\lambda_7\,,
\label{eq:rels}
\eeq
is preserved by renormalization to all orders of perturbation theory, when considering only the scalar and gauge sectors. These relations cannot be reproduced by any of the six global symmetries mentioned above, nor by combinations thereof. We can obtain them, however, using a (seemingly) formal procedure: 
requiring invariance under the transformation $r_0\rightarrow -r_0$, $\vec{r}\rightarrow \vec{r}$ which, if we rewrite the potential as
\beq
V \,=\, M_0\,r_0\,+\,\Lambda_{00}\,r_0^2 \,-\,{\vec{M}}\,. \, \vec{r}
\,-\,2\,\left({\vec{\Lambda}}\,.\,\vec{r}\right)\,r_0 \,+\,
 {\vec{r}\,}\,.\,\left(\Lambda\,\vec{r}\right)\,.
 \label{eq:potbi}
\eeq
implies that $M_0 = 0$ and $\vec{\Lambda} = \vec{0}$ -- and considering the definition of $M_0$ and $\vec{\Lambda}$, this corresponds
exactly to the parameter relations of eq.~\eqref{eq:rels}.
This transformation is the reason for the expression ``$r_0$-symmetries", but it carries an obvious problem: given the definition
of $r_0$ in eq.~\eqref{eq:rs}, there is no unitary transformation involving doublets or their complex conjugates that can achieve
$\Phi_1^\dagger \Phi_1 + \Phi_2^\dagger \Phi_2 \rightarrow -(\Phi_1^\dagger \Phi_1 + \Phi_2^\dagger \Phi_2)$.

In~\cite{Ferreira:2023dke} it was seen that in order to achieve $r_0\rightarrow -r_0$, $\vec{r}\rightarrow \vec{r}$ one would need to
hypothesize that the scalar doublets transform as~\footnote{We chose a transformation different from that used in~\cite{Ferreira:2023dke} 
by an overall minus sign affecting both doublets. This is clearly physically equivalent since the theory is invariant under the
hypercharge group $U(1)$, and is a choice designed for consistency with the notation of ref.~\cite{Ferreira:2010bm}.}
\beq
\begin{array}{rclcrcl}
\Phi_1 &\to& \Phi_2^* & , &
\Phi_1^\dagger &\to& -\Phi_2^T, \\
\Phi_2 &\to& -\Phi_1^* & , &
\Phi_2^\dagger &\to&  \Phi_1^T\,,
\end{array}
\label{eq:trandou}
\eeq
in which we see that each doublet and its hermitian conjugate transform differently than usual: namely, the transformation of the 
hermitian conjugate of $\Phi_1$ ($\Phi_2$) is {\em not} the hermitian conjugate of the transformation of $\Phi_1$ ($\Phi_2$),
exactly the type of generalized field transformations discussed in~\cite{Trautner:2025yxz}, which 
excludes the usual unitary transformations on doublets or their complex conjugates. One way to obtain the strange 
transformations of eq.~\eqref{eq:trandou} is to perform an {\em imaginary scaling} on the scalar doublets' real components. Indeed,
if we parameterize both doublets as
\bea
\Phi_1=\frac{1}{\sqrt{2}}
\begin{pmatrix}
	\phi_1+i\phi_2 \\
	\phi_3+i\phi_4
\end{pmatrix}, \quad
\Phi_2=\frac{1}{\sqrt{2}}
\begin{pmatrix}
	\phi_5+i\phi_6 \\
	\phi_7+i\phi_8
\end{pmatrix}\,.
\eea
eq.~\eqref{eq:trandou}, or equivalently $r_0\rightarrow -r_0$, $\vec{r}\rightarrow \vec{r}$, correspond to the following 
transformations on the real components $\phi_i$:
\beq
\begin{array}{rcrcrcrcrcrcrcr}
 \phi_1 &\to& -i\phi_6 &,&  \phi_2 &\to& -i\phi_5 &,&  \phi_3 &\to& -i\phi_8 &,&  \phi_4 &\to& -i\phi_7, \\
\phi_5 &\to& i\phi_2 &,& \phi_6 &\to& i\phi_1 &,& \phi_7 &\to& i\phi_4 &,& \phi_8 &\to& i\phi_3. 
\end{array}
\label{eq:imss}
\eeq
These are very strange transformations: all that one usually does in theories with real scalar fields is consider
their transformations which involve either proper or improper orthogonal rotations between those fields; in either case
the scalars $\phi_i$ are multiplied by real numbers, not imaginary units such as those present above. But an arguably stranger thing 
is then found when one considers the kinetic terms of the doublets, given by the lagrangean
\beq
{\cal L}_{\Phi K}\,=\, \big(\partial_\mu \Phi_1^\dagger\big)  \big(\partial^\mu \Phi_1\big)\,+\,
\big(\partial_\mu \Phi_2^\dagger\big)  \big(\partial^\mu \Phi_2\big)\,.
\eeq
Considering the field transformations of eq.~\eqref{eq:trandou} or, equivalently, of~\eqref{eq:imss}, one would find ${\cal L}_{\Phi K} \to - {\cal L}_{\Phi K}$, and 
the only way to avoid this sign change of the kinetic terms is to perform an imaginary scaling of the spacetime coordinates themselves, to wit
\beq
x_\mu \,\to\,i\,x_\mu\,.
\label{eq:imsx}
\eeq
Let us further consider gauge interactions. To accurately describe
scalar-gauge interactions it is well known that we should replace the partial derivative $\partial_\mu$ in ${\cal L}_{\Phi K}$ by the covariant
derivative $D_\mu$ for the $SU(2)\times U(1)$ gauge group,
\beq
D_\mu\,=\,\partial_\mu\,+\,i\,\frac{g}{2}\,\sigma_a W^a_\mu\,+\,i\,Y\frac{g^\prime}{2}B_\mu\,=\, \partial_\mu\,+\,\frac{i}{2}\,M^{EW}_\mu
\eeq
with $Y$ is the hypercharge (for our scalar doublets $Y = 1$), $\sigma^a$ ($a = 1, 2, 3$) represent the three Pauli Matrices, 
$W^a_\mu$ and $B_\mu$ the usual gauge fields and for future convenience we define the matrix $M^{EW}_\mu$, 
\beq
M^{EW}_\mu\,=\, \begin{pmatrix}
   g^\prime Y B_\mu \,+\, g W^3_\mu & g(W^1_\mu - i W^2_\mu) \\
   g(W^1_\mu + i W^2_\mu) &  g^\prime Y B_\mu \,-\, g W^3_\mu
\end{pmatrix}\,.
\label{eq:MEWmu}
\eeq
To achieve invariance of ${\cal L}_{\Phi K}$ under the transformations of~\eqref{eq:trandou} with $\partial_\mu \to D_\mu$, 
it is necessary to admit that the gauge fields themselves also suffer an imaginary scaling, 
\beq
B_\mu \to i B_\mu \quad , \quad  
W^1_\mu \to i W^1_\mu \quad ,\quad
W^2_\mu \to -i W^2_\mu \quad,\quad
W^3_\mu \to i W^3_\mu\,,
\label{eq:imsg}
\eeq
under which we see~\eqref{eq:MEWmu} transforms as $M^{EW}_\mu \to i {M^{EW}_\mu}^T$. With this transformation along with those of eqs.~\eqref{eq:trandou}
and~\eqref{eq:imsx} we find that
\beq
\begin{array}{rcrcrcr}
D_\mu\Phi_1 &\rightarrow&  i\left(D_\mu\Phi_2\right)^* &,& \left(D_\mu\Phi_1\right)^\dagger &\rightarrow& -i\left(D_\mu\Phi_2\right)^T,\\
D_\mu\Phi_2 &\rightarrow&  -i\left(D_\mu\Phi_1\right)^* &,& \left(D_\mu\Phi_2\right)^\dagger &\rightarrow& i\left(D_\mu\Phi_1\right)^T
\end{array}
\eeq
which proves the invariance of the gauge-invariant scalar kinetic terms. We could argue that this procedure is altogether arbitrary, but
let us now look at the gauge fields' kinetic terms,
\beq
{\cal L}_{G K}\,=\,-\frac{1}{4}B_{\mu\nu}B^{\mu\nu}\,-\,\frac{1}{4}W^a_{\mu\nu}W^{a\,\mu\nu},
\eeq
where as usual
\beq
B_{\mu\nu}\,=\,\partial_\nu B_\mu \,-\, \partial_\mu B_\nu\;\;\; , \;\;\;
W^a_{\mu\nu}\,=\,\partial_\nu W^a_\mu \,-\, \partial_\mu W^a_\nu \,-\,
g\epsilon^{abc}W^b_\mu W^c_\nu\,.
\eeq
Considering the imaginary scaling of the spacetime coordinates in~\eqref{eq:imsx} and the gauge fields in~\eqref{eq:imsg}, we
find 
\beq
B_{\mu \nu} \to  B_{\mu \nu} \;\; , \;\;
W^1_{\mu \nu}  \to  W^1_{\mu \nu} \;\; , \;\;
W^2_{\mu \nu}  \to  -W^2_{\mu \nu} \;\; , \;\;
W^3_{\mu \nu} \to W^3_{\mu \nu},
\eeq
and therefore ${\cal L}_{G K}$ is invariant under these combined transformations. The impressive aspect of this invariance is the consistency of
the whole procedure: the $r_0$ transformation requires the imaginary scaling of eq.~\eqref{eq:imss} (or equivalently~\eqref{eq:trandou}); invariance 
of the scalar-only kinetic terms then leads to the imaginary scaling of spacetime coordinates, eq.~\eqref{eq:imsx}; introducing gauge interactions with scalars,
their invariance requires an imaginary scaling of gauge fields, eq.~\eqref{eq:imsg}; and then we find, without further assumptions, that the gauge fields'
kinetic terms are also invariant under this set of transformations. Nevertheless, the idea of performing an imaginary scaling on real fields 
and spacetime coordinates is so bizarre that these symmetries -- if symmetries they are -- were dubbed GOOFy in~\cite{Ferreira:2023dke}, taking advantage
of the names of the authors of that paper.

\section{Inclusion of fermions}
\label{sec:yuk}

The procedure outlined above and introduced in~\cite{Ferreira:2023dke} concerned the scalar and gauge sectors only. It was shown in that paper that
the RG-invariance of the parameter relations~\eqref{eq:rels} held to all orders of perturbation theory when just those interactions were taken into
account. An all-order result involving fermions, however, could not be obtained in~\cite{Ferreira:2023dke}, but direct inspection of the $\beta$-functions
of the model confirmed invariance of~\eqref{eq:rels} up to at least two-loops, even when Yukawa interactions were taken into account. This was shown 
to occur for two specific Yukawa textures, those of the so-called CP2 and CP3 versions of the 2HDM~\cite{Ferreira:2010bm}. 
In~\cite{Trautner:2025prm,deBoer:2025jhc} GOOFy-like transformations were suggested for the SM fermions to address issues of hierarchy: 
the resulting theory is a scale-invariant model, without mass terms; this conclusion holds even if one adds vector-like 
quarks~\cite{deBoer:2025jhc}, the imaginary scaling of fermion fields introduced eliminating possible fermionic mass ´terms as weel.

In this section we wish to 
obtain the necessary transformations to impose on the fermion fields of the 2HDM so as to obtain invariance under the GOOFy transformations
that affect scalars, eqs.~\eqref{eq:trandou} or~\eqref{eq:imss}, gauge fields, eq.~\eqref{eq:imsg}, and spacetime coordinates, eq.~\eqref{eq:imsx}, that justify the CP2 and CP3 Yukawa textures found in~\cite{Ferreira:2023dke}. 
We will only consider quarks in this paper, the inclusion of leptons would be treated much in the same manner as we are about to do. The most 
general 2HDM quark Yukawa lagrangian may be written as
\beq
-{\cal L_Y}\,=\, \overline{Q}_L\Gamma_1 \Phi_1 n_R\,+\,\overline{Q}_L\Gamma_2 \Phi_2 n_R\,+\,
\overline{Q}_L\Delta_1 \tilde{\Phi}_1 p_R\,+\,\overline{Q}_L\Delta_2 \tilde{\Phi}_2 p_R
+{\rm h.c.}\,,
\label{eq:yuk}
\eeq
where $\Gamma_i$ and $\Delta_i$ are $3\times 3$ complex matrices containing the Yukawa couplings of the model. The right handed $n_R$/ $p_R$ fields
are 3-vectors in flavour space, containing the negatively/positively charged quarks which, upon rotation to the mass basis, will originate 
the right-handed down/up quarks. Likewise, the left handed $Q_L$ doublets also are also 3-vectors in flavour space, containing the unrotated 
left quark fields. $\tilde{\Phi}_i = i \sigma_2 \Phi_i^*$, as usual, stands for the charge conjugate of the scalar doublets.

Let us now draw inspiration from the $r_0$ transformations we wrote earlier for the scalar doublets, eq.~\eqref{eq:trandou}. If we consider the
transformations hypothesised for $\Phi_1$ and $\Phi_2$ in that equation, we recognize the form of the generalized CP transformation (GCP) applied to
the 2HDM originating the so-called CP2 model~\cite{Maniatis:2007de,Maniatis:2009vp,Ferreira:2010bm}, to wit $\Phi_1 \to \Phi_2^*$ and 
$\Phi_2 \to -\Phi_1^*$. This leads
us to try a similar GCP transformation for the quark fields:
\bea
Q_L &\to &  X_\alpha\,\gamma^0 \,C\,Q_L^* \nonumber \\
n_R &\to &  X_\beta\,\gamma^0\,C\,n_R^* \nonumber  \\
p_R &\to &  X_\gamma\,\gamma^0\,C\,p_R^*
\label{eq:imsq}
\eea
where $C$ is the charge conjugation matrix (obeying $C^2 = -1$, $C^{-1} = C^\dagger = -C$, and $C\gamma^\mu C^{-1} = -{\gamma^\mu}^T$) and, following the notation of~\cite{Ferreira:2010bm}, the three $X_\theta$ are $3\times 3$ unitary matrices describing transformations in
flavour space, mixing the different quarks. If we were to now consider how, for instance, $\bar{n}_R$ would transform, we would have
\beq
\overline{n}_R \,=\, n_R^\dagger \,\gamma^0\,\to\, n_R^T\,C^\dagger \gamma^0 \,X^\dagger_\beta\,\gamma^0\,=\, 
-n_R^T\,C \,X^\dagger_\beta\,.
\eeq
As in the scalar doublets GOOFy transformations, however, we will have the hermitian conjugates of the  
quark fields transforming differently than the hermitian conjugates of the transformations above, namely:
\bea
\overline{Q}_L &\to & -\,\eta_Q\,Q_L^T\,C\, X^\dagger_\alpha \nonumber \\
\overline{n}_R &\to & -\,\eta_n\,n_R^T\,C\, X^\dagger_\beta  \nonumber \\
\overline{p}_R &\to & -\,\eta_p\,p_R^T\,C\, X^\dagger_\gamma\,,
\label{eq:imsqb}
\eea
where the $\eta_x$ are complex phases~\footnote{Different possibilities for these transformations are discussed 
in~\cite{Trautner:2025prm}.}. For any of the fermionic fields $\psi$ 
above the kinetic terms will transform as
\bea
i\,\overline{\psi}\,\slashed{\partial}\,\psi & \to & 
-\,i\, \eta \,\psi^T \,C\,X^\dagger_\theta\,(-i\,\slashed{\partial})\,X_\theta\,\gamma^0 \,C\,\psi^*  \nonumber \\
 & = & -\,\eta\, \psi^T \,C\,\gamma^\mu\,C\,\gamma^0 \,\partial_\mu\, \psi^* 
 \, = \, \eta\, \psi^T \, {\gamma^\mu}^T\,\gamma^0 \,(\partial_\mu\, \psi^*)  \nonumber \\ 
 & = & - \, \eta\, (\partial_\mu \overline{\psi} ) \, \gamma^\mu\, \psi \,=\, 
  \eta\,\overline{\psi} \, \slashed{\partial}\, \psi\,,
 \label{eq:invlk}
\eea
where we used the unitarity of the $X$ matrices, $X_\theta^\dagger X_\theta =  \mathbb{1}$; in the third line a ``-" sign 
appears due to the transposition of a fermionic operator; and in the final step we integrated by parts and dropped a total divergence.
Invariance of the fermionic kinetic terms therefore requires that the $\eta_x$ phases present in the transformations of
eqs.~\eqref{eq:imsqb} obey
\beq
 \eta_Q\,=\,\eta_n\,=\, \eta_p\,=\, i\,.
\label{eq:etas}
\eeq
Adding the gauge interactions of fermions, they are described by modifying the kinetic lagrangian with the appropriate 
covariant derivative, ${\cal L}_{\psi K}\,=\,i\,\overline{\psi}\,\slashed{D}\,\psi$, with
\beq
D_\mu\,=\,\partial_\mu\,+\,\frac{i}{2}\,M_\mu\,=\,\partial_\mu\,+\,\frac{i}{2}\,M^{S}_\mu\,+\,\frac{i}{2}\,M^{EW}_\mu
\eeq
where the electroweak gauge boson contribution $M^{EW}_\mu$ was defined in eq.~\eqref{eq:MEWmu} 
and the gluon contribution is given, in terms of the strong coupling
constant $g_s$, the Gell-Mann matrices $\lambda^a$ and gluon fields $G^a_\mu$, by
\beq
M^{S}_\mu\,=\,  g_s \,\lambda^a\,G^a_\mu\,=\, g_s\, \begin{pmatrix}
   G^3_\mu \,+\, \frac{1}{\sqrt{3}} G^8_\mu & G^1_\mu - i G^2_\mu & G^4_\mu - i G^5_\mu \vspace{2mm} \\ 
   G^1_\mu + i G^2_\mu &  -G^3_\mu \,+\, \frac{1}{\sqrt{3}} G^8_\mu & G^6_\mu - i G^7_\mu \vspace{2mm} \\
   G^4_\mu + i G^5_\mu & G^6_\mu + i G^7_\mu & -\frac{2}{\sqrt{3}} G^8_\mu
\end{pmatrix}\,.
\label{eq:MSmu}
\eeq
We need to know what are the GOOFy transformations of the gluon fields. In analogy with the electroweak ones, they transform as
\beq
\begin{array}{rcrcrcrcrcrcrcr}
G^1_\mu & \to & i G^1_\mu &,& G^2_\mu & \to & - i G^2_\mu &,& G^3_\mu & \to & i G^3_\mu &,& G^4_\mu & \to & i G^4_\mu ,\\
G^5_\mu & \to & -i G^5_\mu &,& G^6_\mu & \to & i G^6_\mu &,& G^7_\mu & \to & -i G^7_\mu &,& G^8_\mu & \to & i G^8_\mu \,\,
\end{array}
\label{eq:imsgs}
\eeq
where the gluon fields appearing in $M^S_\mu$ multiplied by $i$ transform with factors of $-i$, the rest by factors of $i$, so
that, analogously to $M^{EW}_\mu$, this matrix transforms as $M^S_\mu \to i {M^S_\mu}^T$. Thus, the fermion kinetic + gauge terms
transform as
\bea
i\,\overline{\psi}\,\slashed{D}\,\psi & \to & 
\eta \,\left[\overline{\psi} \, \slashed{\partial}\, \psi \,+\, \frac{1}{2}\,
\psi^T \,C\,X^\dagger_\theta\,(i\,M^T_\mu)\,\gamma^\mu\,X_\theta\,\gamma^0 \,C\,\psi^* \right] \nonumber \\
 & = & 
\eta \,\left[\overline{\psi} \, \slashed{\partial}\, \psi \,-\, \frac{i}{2}\,
\psi^T \,M^T_\mu\,{\gamma^\mu}^T\,\gamma^0 \,\psi^* \right] \nonumber 
\,=\,
 \eta \,\left[\overline{\psi} \, \slashed{\partial}\, \psi \,+\, \frac{i}{2}\,
\overline{\psi}  \,M_\mu\,\gamma^\mu\,\psi \right] \,,
\eea
where we used the same manipulations as in eq.~\eqref{eq:invlk} 
and we see that the same condition of eq.~\eqref{eq:etas} which ensured invariance of the kinetic terms, all $\eta = i$, 
is in fact sufficient to do the same for the full gauge invariant kinetic lagrangian. 
Since we had to introduce GOOFy transformations for the gluon fields we must verify whether
they leave their kinetic terms invariant. The corresponding terms in the lagrangian are
\beq
{\cal L}_{G K}\,=\,-\frac{1}{4}\,G^A_{\mu\nu}\,G_{A\mu\nu}\,, \quad \mbox{with}  \quad
G^A_{\mu\nu}\,=\, \partial_\mu G^A_\nu\,-\,\partial_\nu G^A_\mu\,+\,g_s f^{ABC} G^B_\nu G^C_\nu
\label{eq:king}
\eeq
where $f^{ABC}$ are the $SU(3)$ structure constants. It is then easy to see that the transformations~\eqref{eq:imsgs}, along
with~\eqref{eq:imsx}, affect the gluon gauge tensor as
\beq
\begin{array}{rcrcrcrcrcrcrcr}
G^1_{\mu\nu} & \to & G^1_{\mu\nu} &,& G^2_{\mu\nu} & \to & - G^2_{\mu\nu} &,& 
G^3_{\mu\nu} & \to & G^3_{\mu\nu} &,& G^4_{\mu\nu} & \to & G^4_{\mu\nu} \vspace{2mm}  \\
G^5_{\mu\nu} & \to & -G^5_{\mu\nu} &,& G^6_{\mu\nu} & \to & G^6_{\mu\nu} &,& 
G^7_{\mu\nu} & \to & -G^7_{\mu\nu} &,& G^8_{\mu\nu} & \to & G^8_{\mu\nu}
\end{array}
\eeq
and therefore the kinetic terms in eq.~\eqref{eq:king} are indeed left invariant.

Finally, the Yukawa terms: let us show how the scalar and fermionic GOOFy transformations of eqs.~\eqref{eq:imss},~\eqref{eq:imsq}
and~\eqref{eq:imsqb} leave the terms concerning the down quarks invariant, the calculation for the up quark terms follows in
the same manner. We start from eq.~\eqref{eq:yuk}, writing explicitly the hermitian conjugate term, and proceed from there:
\bea
-{\cal L}_{Yd} & = & \overline{Q}_L\,\left(\Gamma_1 \Phi_1 \,+\,\Gamma_2 \Phi_2\right)\,n_R\,+\,
\overline{n}_R\,\left(\Gamma_1^\dagger \Phi_1^\dagger \,+\,\Gamma_2^\dagger \Phi_2^\dagger\right)\,Q_L\; \to 
\label{eq:yuko} \\
 & \to &  -\,\eta_Q \, Q^T_L\,C\,X_\alpha^\dagger\,\left( \Gamma_1 \Phi_2^* - \Gamma_2 \Phi_1^*\right)\, X_\beta\,\gamma^0\,C\,n^*_R
 \nonumber \\
 & &  -\,
 \eta_n \, n^T_R\,C\,X_\beta^\dagger\,\left(-\Gamma_1^\dagger \Phi_2^T + \Gamma_2^\dagger \Phi_1^T\right)\, X_\alpha\,\gamma^0\,C\,Q^*_L 
  \nonumber \\
  & = & \,i\, \overline{Q}_L\,X_\alpha^T\,\left(\Gamma_2^* \Phi_1 - \Gamma_1^* \Phi_2 \right)\, X_\beta^*\,n_R
\label{eq:yukf1} \\
  & &  
 +i\,\overline{n}_R\,X_\beta^T\,\left(- \Gamma_2^T \Phi_1^\dagger + \Gamma_1^T \Phi_2^\dagger \right)\, X_\alpha^*\,Q_L
 \,,
\label{eq:yukf2}
\eea
where we used $\eta_Q = \eta_n = i$ and in the final step we again transposed all terms, which got rid of an overall minus sign. 
We therefore find that the lagrangian is invariant under these 
transformations if and only if the Yukawa matrices obey certain conditions -- equaling the  first term in~\eqref{eq:yuko} with~\eqref{eq:yukf1} and performing
a complex conjugation, we obtain
\bea
 \Gamma_1^* &=& -  i\,X_\alpha^\dagger \,\Gamma_2\,X_\beta \,\nonumber \\
 \Gamma_2^* &=& \;\;\;i\, X_\alpha^\dagger \,\Gamma_1\,X_\beta
\label{eq:gamma}
\eea
and equalling the second term in~\eqref{eq:yuko} with~\eqref{eq:yukf1} would yield exactly the same equations. We thus find that the 
$\Gamma_i$ matrices obey, other than the factor of ``$i$" in the equations above, the same conditions as those of the CP2 model 
(see eq. (8) of~\cite{Ferreira:2010bm}). But if we redefine the Yukawa matrices as $\overline{\Gamma}_i = e^{i\pi/4}\Gamma_i$, we 
see that the $\overline{\Gamma}$ obey exactly the CP2 symmetry conditions, which imply a very restrictive Yukawa texture.
Going back to $\Gamma_i$, then, one obtains~\footnote{We are following the same procedure as in reference~\cite{Ferreira:2010bm}
to obtain the Yukawa textures dictated by the symmetry conditions imposed by eq.~\eqref{eq:gamma}; namely, we are
assuming that the results of~\cite{Ecker:1987qp} apply to the GOOFy transformations considered here, and it is always
possible to find a quark basis such that the $X_{\alpha,\beta}$ matrices assume a very simple form, in which they 
become orthogonal and depend on a single rotation angle.}
\beq
\Gamma_1 \,=\,\begin{pmatrix} a_{11} & a_{12} & 0 \\ a_{12} & -a_{11} & 0 \\
0 & 0 & 0 \end{pmatrix}\;\;,\;\;
\Gamma_2 \,=\,\begin{pmatrix} -a_{12}^* & a_{11}^* & 0 \\ a_{11}^* & a_{12}^* & 0 \\
0 & 0 & 0 \end{pmatrix}\;.
\label{eq:yukcp2}
\eeq
where the $a_{ij}$ are complex coefficients in a specific basis of quark fields, and we absorbed an overall $e^{i\pi/4}$
phase in their definition. Another way of achieving this is to perform a basis change on the right-handed quark
fields to absorb the ``i" factor in eq.~\eqref{eq:gamma}. An analogous expression (with different 
complex coefficients $b_{ij}$) is found for the up-quark matrices $\Delta_i$. We therefore conclude that the $r_0$-symmetry
extended to the fermion sector requires CP2-like Yukawa matrices, which explains the result obtained in Section 5
ref.~\cite{Ferreira:2023dke} via direct inspection of $\beta$-functions up to two-loops - it was argued in that reference that
CP2 Yukawa textures would preserve the symmetry relations of the dimensionless parameters imposed by the $r_0$-symmetry,
since they were indistinguishable from the CP2-symmetry ones, but invariance of the dimensional parameters' relations
could not be proved there, only verified up to two-loops. The present calculation, however, shows that CP2 Yukawa 
structures are required under the imaginary scaling of all fields that leads to eqs.~\eqref{eq:gamma}.

In reference~\cite{Ferreira:2023dke} it was possible to prove RG invariance to all orders when only the scalar and gauge sectors
were considered. It would be useful to be able to use the same argument including fermion contributions as well.
To begin, let us consider how the Yukawa sector behaves regarding scalar basis changes. The work of~\cite{Sartore:2022sxh} casts 
these fermion basis transformations in bilinears involving the Yukawa matrices, showing for instance that the quantity
\beq
Y^0_d \,=\, \Gamma_1\,\Gamma_1^\dagger \,+\, \Gamma_2\,\Gamma_2^\dagger
\eeq
is basis-invariant, whereas, in terms of the  Pauli matrices $\vec{\sigma} = (\sigma_1,\sigma_2,\sigma_3)$,
\beq
\vec{Y}_d\,=\, \Gamma_a\,(\vec{\sigma})_{ab} \,\Gamma_b \,=\,\begin{pmatrix}
    \Gamma_1\,\Gamma_2^\dagger \,+\, \Gamma_2\,\Gamma_1^\dagger \\ 
    i\,(\Gamma_1\,\Gamma_2^\dagger \,-\, \Gamma_2\,\Gamma_1^\dagger ) \\
    \Gamma_1\,\Gamma_1^\dagger \,-\, \Gamma_2\,\Gamma_2^\dagger
\end{pmatrix}
\label{eq:vecY}
\eeq
behaves under scalar basis changes like a vector, much like $\vec{M}$ or $\vec{\Lambda}$. 
Similar structures may be built with the up quark Yukawa matrices $\Delta$, and indeed we can conceive of an arbitrary number
of Yukawa matrices combined in this manner to make analogous objects. 

We can thus follow the arguments of~\cite{Bednyakov:2018cmx} and argue that the  all-orders $\beta$-functions for the parameter $M_0$ 
can be expressed as
\beq
\beta_{M_0}\,=\, b_0 \,M_0\,+\,b_1\,\vec{\Lambda} \cdot\vec{M}\,+\, b_2 \,\vec{\Lambda} \cdot\left(\Lambda \vec{M}\right)
\,+\, b_3 \,\vec{\Lambda} \cdot\left(\Lambda^2 \vec{M}\right)\,+\,\,  b_4\,\vec{Y} \cdot\vec{M}\,,
\label{eq:betaM0}
\eeq
with $\vec{Y}$ a vector formed with Yukawa matrices like that of eq.~\eqref{eq:vecY} (there will be many such possible vectors), 
and 
with basis-invariant coefficients $b_i$ depending on dimensionless couplings of the model, and $\vec{\Lambda}$, 
$\vec{M}$ and the matrix $\Lambda$ defined in eqs.~\eqref{eq:defbil} and~\eqref{eq:Lambda}. 
Basis-invariant quantities built from Yukawa couplings, such as $Y^0_{u,d}$ or $\vec{Y}_d \cdot \vec{Y}_d$ or many such others,
will be contained in the several basis-invariant coefficients $b_{1 \dots 4}$.

The reasoning behind eq.~\eqref{eq:betaM0}
is simple: since $M_0$ is a basis-invariant quantity so should its $\beta$-function be, as well as have the same dimensions of
mass. All these contributions can be understood in this manner (see~\cite{Bednyakov:2018cmx} and~\cite{Ferreira:2023dke} 
for a detailed explanation of the scalar/gauge contributions). So the Yukawa contributions to the all-orders
$M_0$ $\beta$-function end up being formally simple, but unfortunately it is difficult to achieve much with them: for the $r_0$-symmetry,
since $\vec{M}\neq \vec{0}$, one would need to argue that due to the symmetry being imposed on the fermionic sector all vectors
$\vec{Y}$ would necessarily be zero, and it is not clear why that should obviously hold to all orders. For instance, while the CP2
Yukawa structures of eq.~\eqref{eq:yukcp2} clearly yield $\vec{Y}_d = \vec{Y}_u = \vec{0}$, that does not imply that any
such vectors built with products of an arbitrary number of $\Gamma$ and/or $\Delta$ matrices should obey the same property. 
Thus an all-order result
for RG invariance of $M_0 = 0$ including Yukawa interactions remains elusive.

\section{Other possibilities}
\label{sec:omod}

Having proven that the CP2 Yukawa textures are obtained in a consistent way in the imaginary scaling of spacetime coordinates 
and fields described in the previous section, we will now investigate how other GOOFy symmetries might be extended to the fermion 
sector. 

\subsection{GOOFy GCP symmetries}
\label{sec:ggcp}

An obvious starting point is going back to the scalar transformations of the $r_0$-symmetry, eq.~\eqref{eq:trandou},
which we already mentioned that are very similar to those of the CP2 2HDM -- except for the fact that each doublet and its conjugate transform
in different ways. The CP2 symmetry, as shown in refs.~\cite{Ivanov:2007de,Ferreira:2010bm}, is a special case of a generalized CP (GCP)  symmetry,
obtained via a transformation of the type $\Phi_i \rightarrow X_{ij} \Phi_j^*$, with an implicit sum on the index $j$ and
a  choice of scalar basis alway being possible so that the unitary $2\times 2$ matrix $X$ is expressed as
\beq
X\,=\,\begin{pmatrix}
    \cos\theta & \sin\theta \\
    - \sin\theta & \cos\theta
\end{pmatrix}\,,
\eeq
with $\theta \in [0\,,\,\pi/2]$ without loss of generality. In this notation, then, CP2 corresponds to $\theta = \pi/2$; the CP1 symmetry is the 
``usual" CP symmetry,
with $\theta = 0$ and corresponding to the field transformations  $\Phi_i \rightarrow \Phi_i^*$; and CP3 would correspond to all other values of $\theta$
in $]0\,,\,\pi/2[$. An obvious generalization of the $r_0$-transformation of eq.~\eqref{eq:trandou} is therefore
\beq
\begin{array}{rclcrcl}
\Phi_1 &\to& \cos\theta \,\Phi_1^* + \sin\theta\,\Phi_2^* & , &
\Phi_1^\dagger &\to& -\cos\theta \,\Phi_1^T - \sin\theta\,\Phi_2^T, \\
\Phi_2 &\to& -\sin\theta \,\Phi_1^* + \cos\theta\,\Phi_2^* & , &
\Phi_2^\dagger &\to&  \sin\theta \,\Phi_1^T - \cos\theta\,\Phi_2^T\,,
\end{array}
\label{eq:r0gen}
\eeq
from which one very quickly gets
\bea
r_0 &\rightarrow & -\,r_0\,, \nonumber \\
r_1 &\rightarrow & -\cos(2\theta)\,r_1\,+\,\sin(2\theta)\,r_3 \,, \nonumber \\
r_2 &\rightarrow & r_2\,, \nonumber \\
r_3 &\rightarrow & -\sin(2\theta)\,r_1\,-\,\cos(2\theta)\,r_3
\eea
with the initial $r_0$-symmetry corresponding to $\theta = \pi/2$. Following all the steps in the calculation that led to 
eq.~\eqref{eq:yukf2}, then, we are left with
\bea
-{\cal L}_{Yd} & = & 
 \,i\, \overline{Q}_L\,X_\alpha^T\,\left[(-\cos\theta \,\Gamma_1^* + \sin\theta \,\Gamma_2^* )\Phi_1 + (-\sin\theta\,\Gamma_1^* - 
 \cos\theta\,\Gamma_2^*) \Phi_2 \right]\, X_\beta^*\,n_R \nonumber  \\
  & &  
 +i\,\overline{n}_R\,X_\beta^T\,
 \left[(\cos\theta \,\Gamma_1^T - \sin\theta \,\Gamma_2^T )\Phi_1^\dagger + (\sin\theta\,\Gamma_1^T + 
 \cos\theta\,\Gamma_2^T) \Phi_2^\dagger \right]\, X_\alpha^*\,Q_L
 \,,
\label{eq:yukrog}
\eea
and therefore, absorbing a factor of ``-i" in the $n_R$ fields through a basis redefinition, we are left with the equations that the $\Gamma$ matrices 
must obey,
\bea
\Gamma_1^* &=& X_\alpha^\dagger\,\left( \cos\theta \,\Gamma_1 \,-\, \sin\theta \,\Gamma_2  \right)\,X_\beta\,, \nonumber \\
\Gamma_2^* &=& X_\alpha^\dagger\,\left( \sin\theta \,\Gamma_1 \,+\, \cos\theta \,\Gamma_2  \right)\,X_\beta\,,
\label{eq:yukgcp}
\eea
which are {\em exactly} the conditions for invariance under GCP symmetries found in ref.~\cite{Ferreira:2010bm} (see eq. (8) in that paper).
An analogous set of equations is found for the up-quark $\Delta$ Yukawa matrices.  
An appropriate basis choice for the quark fields reduces the $3\times 3$ unitary matrices $X_x$ to rotation matrices depending on a single
angle (see~\cite{Ferreira:2010bm} for details), and the resulting Yukawa textures are quite simple. For $\theta = \pi/2$ we have already seen 
the result in eq.~\eqref{eq:yukcp2}. Two other interesting cases may be obtained within the remaining angles.

\subsubsection{The $\theta = \pi/3$ case}

As shown in ref.~\cite{Ferreira:2010bm}, the only angle $0 < \theta< \pi/2$ that yields six massive quarks is $\theta = \pi/3$, for which
eqs.~\eqref{eq:yukgcp} yield
\beq
\Gamma_1 \,=\,\begin{pmatrix} i\,a_{11} & i\,a_{12} & a_{13} \\ i\,a_{12} & -i\,a_{11} & a_{23} \\
a_{31} & a_{32} & 0 \end{pmatrix}\;\;,\;\;
\Gamma_2 \,=\,\begin{pmatrix} i\,a_{12} & -i\,a_{11} & -a_{23} \\ -i\,a_{11} & -i\,a_{12} & a_{13} \\
-a_{32} & a_{31} & 0 \end{pmatrix}\,,
\label{eq:yukcp3}
\eeq
with the coefficients $a_{ij}$ real, and analogous Yukawa textures for the $\Delta$ matrices, with real $b_{ij}$ coefficients. Any other
choice of $\theta$ in this interval yields a massless down and up quark (indeed, the $\theta = \pi/2$ case, eq.~\eqref{eq:yukcp2}, has 
the same pathology). This CP3 Yukawa sector predicts six massive
quarks (unlike the CP2 case) and does fit the magnitudes of the CKM matrix but, as was discussed in ref.~\cite{Ferreira:2010bm},
fails to account for the correct value of the Jarlskog invariant.
Regarding the scalar sector, it can easily be seen that, besides the conditions of eq.~\eqref{eq:rels}, the parameters 
of the potential now also  must obey
\beq
m^2_{11} - m^2_{22}\,=\,\mbox{Re}(m^2_{12})\;=\;0\;\;\; , \;\;\; \mbox{Im}(\lambda_5) = 0\;\;\; , \;\;\;\lambda_6 = \lambda_7 = 0 \;\;,\;\; 
\lambda_5 = \lambda_1 - \lambda_3 - \lambda_4\,.
\label{eq:relsCP3}
\eeq
These match the ``0CP3" case of ref.~\cite{Ferreira:2023dke} for the quartic couplings, though unlike in that case they leave open the possibility of 
$\mbox{Im}(m^2_{12})\neq 0$. 

Notice that this symmetry forces $M_0 = m^2_{11} + m^2_{22} = 0$ and, separately, $M_1 = \mbox{Re}(m^2_{12}) = 0$ and
$M_3 = m^2_{22} - m^2_{11} = 0$ --
the Yukawa couplings of eq.~\eqref{eq:yukcp3} must preserve, under RG running, each of these conditions separately as well. 
This then explains the result
observed up to two-loops in~\cite{Ferreira:2023dke}, that these Yukawa structures left intact, under renormalization, 
the condition $m^2_{11} + m^2_{22} = 0$ 
for a model with that same quartic couplings but non-zero $m^2_{11}$, $m^2_{22}$ and $m^2_{12}$ -- such a theory is a 
softly broken version of the model
with parameters given by~\eqref{eq:rels} and~\eqref{eq:relsCP3}. As such,
we should expect that $\beta_{m^2_{11} + m^2_{22}} = 0$ still holds, even after relaxing the conditions $M_1 = 0$ and $M_3 = 0$, since 
the condition $m^2_{11} + m^2_{22} = 0$  is a leftover symmetry of the model even after the soft breaking chosen -- in fact, the model is left
with an $r_0$ symmetry, so that the scalar potential obeys the restrictions in eq.~\eqref{eq:rels}. The dimensionless couplings' running then
will preserve the leftover conditions of the original model, to wit $m^2_{11} + m^2_{22} = 0$. 

\subsubsection{The $\theta = 0$ case: GOOFy CP1 model}

The case $\theta = 0$ is also interesting: it leads to $r_0\rightarrow -r_0$, $r_1\rightarrow -r_1$, $r_2\rightarrow r_2$ and $r_3\rightarrow -r_3$. 
This corresponds to a different imaginary scaling than the one of the original $r_0$ symmetry, eq.~\eqref{eq:imss}: for instance, the upper two 
real components of $\Phi_1$ transform as
\beq
\phi_1\,\rightarrow\,-i\,\phi_2\;\;\; , \;\;\; \phi_2\,\rightarrow\,-i\,\phi_1\,,
\eeq
with analogous transformations for other pairs of $\phi_i$. The constraints on the scalar sector then become
\beq
m^2_{11} \,=\, m^2_{22}\,=\,\mbox{Re}(m^2_{12})\;=\;0\;\;\; , \;\;\; \mbox{All quartic couplings real}\,.
\label{eq:relsCP0}
\eeq
Notice that these are not the same constraints as seen in the $r_0$-symmetric potentials studied 
in~\cite{Ferreira:2023dke} -- the quartic
couplings, in particular, are very different from those shown in eq.~\eqref{eq:rels}. This is indeed yet another possible 
symmetry for the
2HDM, a GOOFy CP1 model (called ``$CP1_G$" in~\cite{Trautner:2025yxz}, though its fermion sector was not treated in that work), 
and one we can extend to the fermion sector with ease:  from eq.~\eqref{eq:r0gen} with $\theta = 0$, 
we see that the Yukawa matrices obey
\beq
\Gamma_1^*\,=\,\Gamma_1\;\;\; , \;\;\; \Gamma_2^*\,=\,\Gamma_2\,,
\eeq
for the simplest case where we choose the unitary matrices $X_\alpha$ and $X_\beta$ equal to the identity matrix. Therefore the Yukawa matrices are
found to be real, though the model is explicitly CP-breaking, since the $m^2_{12}$ term is proportional to the imaginary unit $i$, and that phase cannot
be absorbed by any field redefinition. A few observations can be made about the phenomenology of this model:
\begin{itemize}
\item 
Spontaneous symmetry breaking is possible, with neutral vacuum expectation values given by  $\langle \Phi_1 \rangle = v_1/\sqrt{2}$
and $\langle \Phi_2 \rangle = v_2 \,e^{i\delta}/\sqrt{2}$, with $v_1^2 + v_2^2 = v^2 = (246\mbox{ GeV})^2$. Using the
notation $m^2_{12} \equiv \mbox{Im}(m^2_{12})$, the vev phase $\delta$ may be determined by one (of three) of the non-trivial 
minimization equations,
\beq
m^2_{12}\,\cos\delta \,-\,\frac{1}{2}\,\left[\lambda_6 v_1^2 + \lambda_7 v_2^2 + 2\lambda_5 v_1 v_2 \cos 2\delta
\right]\,\sin\delta\,=\,0\,.
\eeq
\item Since there is only one quadratic parameter in the scalar potential, this model will not have a decoupling limit
and all squared  scalar masses (neutral states $h_{1,2,3}$ or charged $H^\pm$) will be of the order $\lambda_i\,v^2$.
Indeed, a simple calculation leads to
\bea
m^2_{H^\pm} &=& \frac{v^2}{4 (1 + \cos^2\delta) v_1^2 v_2^2}\,\left\{ \frac{}{} \lambda_1 v_1^4 + \lambda_2 v_2^4 + 
\left[ (\lambda_3 + \lambda_4) \sin^2\delta - \lambda_5 \cos^2\delta\right] v_1^2 v_2^2 \right\} \nonumber \\
\sum_{i=1}^3 m^2_{h_i} &=&  4\,m^2_{H^\pm}\,-\,(\lambda_3 - \lambda_4)\,v^2\,,
%
\eea
where we have used the minimization equations to eliminate the $m^2_{12}$, $\lambda_6$ and $\lambda_7$ parameters.
Since the quartic couplings are limited by unitarity constraints~\cite{Ginzburg:2005dt,Kanemura:2015ska}, a rough upper bound of $\sim 800$ GeV
is expected for all scalars of the model, as was observed for other GOOFy models studied in~\cite{Ferreira:2023dke}.
\item A complex CKM matrix is generated in the Yukawa sector -- the quark Yukawa matrices are real due to the symmetry imposed
on the lagrangean, but the vev phase $\delta$ leads to complex quark mass matrices. For instance, the down quark mass
matrix is
\beq
M_d\,=\,\frac{1}{\sqrt{2}}\,(\Gamma_1 \,v_1\,+\, \Gamma_2\,v_2\,e^{i\delta})\,,
\eeq
and likewise for the up quarks. Diagonalization of these generically complex matrices easily yields a complex CKM matrix~\footnote{Of 
course, the parameters of the Yukawa matrices must be chosen to correctly reproduce quark masses and  CKM entries, but remember that each
of those matrices has 9 independent real parameters.}.
\item It is easy to see how the real Yukawa matrices imposed by the GOOFy symmetry lead to RG invariance of all the lagrangean: being real,
they will never generate, at higher orders, complex phases for the scalar quartic couplings or for themselves; those scalar couplings being 
themselves real, they will never lead to, at any order, complex phases in the Yukawa couplings or anywhere in the model.  
Finally, we can show that the conditions on the quadratic couplings are RG-invariant to all orders, at least in the scalar and gauge sectors. 
The all orders $\beta$-function for $M_0 = m^2_{11} + m^2_{22}$ is given by eq.~\eqref{eq:betaM0}
with basis-invariant coefficients $b_i$ depending on the dimensionless couplings (gauge, scalar) of the model, and $\vec{\Lambda}$, 
$\vec{M}$ and the matrix $\Lambda$ defined in eqs.~\eqref{eq:defbil} and~\eqref{eq:Lambda}. The Yukawa-dependent vectors $\vec{Y}$
are analogous to that of eq.~\eqref{eq:vecY}. With the GOOFy conditions
of eq.~\eqref{eq:relsCP0}, the form of these vectors and matrices is
\beq
\vec{M}\,=\,\begin{pmatrix} 0 \\ \times \\ 0  \end{pmatrix} \;\; , \;\; 
\vec{\Lambda}\,=\,\begin{pmatrix} \times \\ 0 \\  \times \end{pmatrix} \;\; , \;\;
\vec{Y}\,=\,\begin{pmatrix} \times \\ 0 \\  \times \end{pmatrix} \;\; , \;\;
\Lambda \,=\,\begin{pmatrix} \times & 0 & \times \\ 0 & \times & 0 \\ \times & 0 & \times \end{pmatrix}\,.
\label{eq:mat0}
\eeq
The only non-obvious structure is $\vec{Y}$, but it is easy to see that the second entry of these vectors is
the imaginary component of a product of several Yukawa matrices -- and if all of them are real, then this
entry in any vector $\vec{Y}$ that we can build with those matrices is bound to be zero. Therefore, we find that
that all the internal products in eq.~\eqref{eq:betaM0} are zero, and therefore $M_0 = 0$ becomes
an all-order fixed point of the RG running of this quantity. Since $\Lambda^n \vec{M} $ // $\vec{M}$, we can write the 
$\beta$-function for $\vec{M}$ in a  simplified manner,
\bea
\beta_{\vec{M}} \,=\, c_0 \,\vec{M}\,+\,c_1^m (M)\,\Lambda^m \vec{\Lambda}\,+\,c^n_2(M)\,\Lambda^n \vec{Y}
\label{eq:betaM}
\eea
with an implicit sum on the indices $m$ and $n$ and  basis invariant coefficients $c_i$, where $c_{0}$  is built exclusively with dimensionless 
couplings and the $c^n_{1,2}(M)$ also include the $M$ parameters. We also observe that for this model one has 
\beq\vec{\Lambda}\cdot \vec{M} \,=\, \vec{Y}\cdot \vec{M} \,=\, 
\vec{\Lambda}\cdot \Lambda^m\,\vec{M} \,=\, \vec{Y}\cdot  \Lambda^n\,\vec{M} \,=\,0\,
\eeq
Then, given that to all orders $M_0 = 0$, the coefficients $c^m_1(M)$ and $c^n_2(M)$ should depend only on scalars built with 
$\vec{M}$ -- but all such scalars
should result from inner products with $\vec{\Lambda}$ or $\vec{Y}$ which, as we have just observed, are zero. We conclude
that all coefficients $c_1^m (M)$ and $c^n_2(M)$ are necessarily zero. 
Therefore, to all orders, $\beta_{\vec{M}}$ // $\vec{M}$ and thus  the form of this vector is preserved by renormalization, to an 
arbitrary number of loops. 
\end{itemize}
We therefore obtained what seems like a phenomenologically viable GOOFy model, with a fermion sector including six massive quarks 
and {\em a priori} a correct
CKM matrix, and a scalar sector with sufficient parameter freedom to fit current constraints from Higgs precision physics and extra
scalars' searches. Not having a decoupling limit, the model can in principle be excluded -- or indeed verified -- by updated LHC searches in 
the next few years. We leave a detailed analysis of its phenomenology (which must include eletric dipole moment constraints and bounds on 
the CP propertes of the Higgs -- see~\cite{Roussy:2022cmp,CMS:2021sdq,ATLAS:2022akr} for experimental results 
and~\cite{Jung:2013hka,Fontes:2017zfn,Abe:2013qla,Cesarotti:2018huy,Fuyuto:2019svr} for 2HDM applications of these constraints) for future work.

\subsection{GOOFy Higgs-family symmetries}
\label{sec:ghf}

Thus far we have been dealing with GOOFy-like field transformations coupled with GCP ones, {\em i.e.} fields being transformed into linear
combinations of its complex conjugates. Another possibility if to couple GOOFy transfoomations with others that transform the fields into linear
combinations of themselves, so-called Higgs-family (HF) transformations. There are three known HF ``regular" symmetries, which will serve as inspiration
for the following sections. 

\subsubsection{The $Z_2$ case}
\label{sec:gz2}

The $Z_2$ symmetric-2HDM is the most studied version of the model, wherein one of the doublets flips sign under the symmetry, the other
remaining untouched. Drawing from our experience in previous sections, we propose a ``GOOFy $Z_2$" symmetry, in which the doublets and their
conjugates transform independently, so that
\beq
\begin{array}{rclcrcl}
\Phi_1 &\to& \Phi_1 & , &
\Phi_1^\dagger &\to& -\Phi_1^\dagger, \\
\Phi_2 &\to& -\Phi_2 & , &
\Phi_2^\dagger &\to&  \Phi_2^\dagger\,,
\end{array}
\label{eq:gz2}
\eeq
These imply $r_0\rightarrow -r_0$, $r_{1,2} \rightarrow r_{1,2}$, $r_3\rightarrow -r_3$, and the real components $\phi_i$ would be 
transformed as
\beq
\begin{array}{rcrcrcrcrcrcrcr}
 \phi_1 &\to& i\phi_2 &,&  \phi_2 &\to& -i\phi_1 &,&  \phi_3 &\to& i\phi_4 &,&  \phi_4 &\to& -i\phi_3, \\
\phi_5 &\to& -i\phi_6 &,& \phi_6 &\to& i\phi_5 &,& \phi_7 &\to& -i\phi_8 &,& \phi_8 &\to& i\phi_7. 
\end{array}
\eeq
Requiring invariance under these transformations requires that the scalar potential parameters obey
\beq
m^2_{11} \,=\, m^2_{22}\;=\;0\;\;\; , \;\;\; \lambda_6\,=\,\lambda_7\;=\;0\,.
\label{eq:relsGZ2}
\eeq
In other words, the impact of~\eqref{eq:gz2} in the quartic couplings is the same of the $Z_2$ symmetry, but in the quadratic
parameters it is reversed: whereas $Z_2$ forces $m^2_{12} = 0$ and leaves $m^2_{11}$ and $m^2_{22}$ untouched, now the 
opposite happens. It can easily be verified that this is not simply a basis change from the $Z_2$-symmetric case, nor does
it correspond to the $0Z_2$ case considered in~\cite{Ferreira:2023dke} (notice how there is no condition $\lambda_1 = \lambda_2$ in
this symmetry). To extend this symmetry to the fermion sector requires that once again fields and their conjugates
transform independently, and we can easily obtain the familiar flavour-conserving Yukawa sectors typical of the $Z_2$ 2HDM.
For instance, requiring invariance of the left quark doublets and right-handed down quark kinetic terms
may be achieved by demanding that those fields transform as
\beq
\begin{array}{rclcrcl}
n_R &\to& -\,n_R & , &
\overline{n}_R &\to& -\,i\,\overline{n}_R, \\
Q_L &\to& i\,Q_L & , &
\overline{Q}_L &\to& \overline{Q}_L\,,
\end{array}
\label{eq:gqz2}
\eeq
and these transformations yield Type-I Yukawa interactions, wherein only the doublet $\Phi_2$ couples to down-quarks. Similar
transformations may be imposed on the up quarks and leptons and will yield scalar-fermion interactions as those of models Type-I, 
or II, etc. All-order RG invariance of the dimensionless couplings is therefore assured. As for the quadratic couplings, for this
symmetry the vectors $\vec{M}$, $\vec{\Lambda}$, $\vec{Y}$ and matrix $\Lambda$ are now of the form
\beq
\vec{M}\,=\,\begin{pmatrix} \times \\ \times \\ 0   \end{pmatrix} \;\; , \;\; 
\vec{\Lambda}\,=\,\begin{pmatrix} 0 \\ 0 \\  \times \end{pmatrix} \;\; , \;\;
\vec{Y}\,=\,\begin{pmatrix} 0 \\ 0 \\  \times \end{pmatrix} \;\; , \;\;
\Lambda \,=\,\begin{pmatrix} \times & \times & 0 \\ \times & \times & 0 \\ 0 & 0 & \times \end{pmatrix}\,.
\label{eq:matgz2}
\eeq
In particular, given that there are no matrices $\Gamma_1$ and $\Delta_1$ in this model, it is quite clear that 
all vectors $\vec{Y}$ that one can build with Yukawa matrices will have the form above. 
Introducing these vectors into eqs.~\eqref{eq:betaM0} and~\eqref{eq:betaM}, we see that for this model $\beta_{M_0} = 0$ 
and  $\beta_{\vec{M}}$ // $\vec{M}$, and thus an all-order RG invariance of the quadratic terms holds.
This is another viable model, with some interesting features: (a) no decoupling limit, once again; (b) explicit
CP breaking, due to the arbitrary phases present in the parameters $m^2_{12}$ and $\lambda_5$;  (c) and no FCNC interactions. 
The model is therefore a particular case of the C2HDM proposed in~\cite{Ginzburg:2002wt} 
(see~\cite{Khater:2003wq,Grzadkowski:2009iz,Arhrib:2010ju,Barroso:2012wz,Fontes:2017zfn} for further work on the C2HDM). 
With spontaneous symmetry breaking occurring with vevs such as $\langle \Phi_1 \rangle = v_1/\sqrt{2}$
and $\langle \Phi_2 \rangle = v_2 \,e^{i\theta}/\sqrt{2}$, with $v_1^2 + v_2^2 = v^2 = (246\mbox{ GeV})^2$, the minimisation
conditions yield several interesting aspects. For instance, if $\tan\beta = v_2/v_1$ as usual, and
 working on a basis where $\lambda_5$ is real, one finds that in this model
\beq
\tan\beta \, = \, \sqrt[4]{\frac{\lambda_1}{\lambda_2}}\,,
\label{eq:tanb} 
\eeq
This is the simplest expression for $\tan\beta$ that the author has encountered. It is worth pointing out 
that the factor involving $\lambda_2$ and $\lambda_1$ also appears in the discriminant which establishes whether, in the (real) 
softly broken 2HDM with a $Z_2$ symmetry,
a given minimum is the global one: it was found in~\cite{Barroso:2013awa}, applying the results from~\cite{Ivanov:2007de},
that a given minimum with vevs $v_1$, $v_2$, was the global one if $D > 0$~\footnote{An analogous condition was deduced
for the most general 2HDM in~\cite{Ivanov:2015nea}.}, with
\beq
D\,=\,m^2_{12}\,\left[m^2_{11} \,-\,\sqrt{\frac{\lambda_1}{\lambda_2}}\,m^2_{22}\right]
\,\left[ \tan\beta \,-\, \sqrt[4]{\frac{\lambda_1}{\lambda_2}} \right]
\eeq
and as we see, the same factor of eq.~\eqref{eq:tanb} appears here. 
To ascertain the non-existence of a decoupling limit we look at the expressions for the scalar masses in this model. We obtain, for the charged one
\beq
m^2_{H^\pm}\,=\,\frac{1}{2}\,\left(\frac{\lambda_1}{\tan^2\beta}\,+\, \lambda_3\right)\,v^2\,.
\eeq
The neutral mass matrix includes the massless Goldstone boson and three neutral scalars, $h_1$, $h_2$ and $h_3$,
which are CP admixtures. The trace of that matrix yields
\beq
m^2_{h_1} + m^2_{h_2} + m^2_{h_3} \,=\, \left(2\frac{\lambda_1}{\tan^2\beta}\,+\, \lambda_3\,+\, \lambda_4\right)\,v^2\,.
\eeq
These two expressions show therefore that no decoupling limit may occur - all masses are combinations of quartic 
couplings times $v$, and unitarity forbids the $\lambda$'s from being too large. A dedicated study is necessary to
verify if this model is excluded by existing data. The fact that no decoupling occurs means that one will need to
make sure the extra scalars are not ruled out by existing LHC searches; and the existence of explicit CP breaking
indicates that bounds from electric dipole moment measurements will severely constrain the model's parameter space.
We leave a thorough study of the model's phenomenology for a follow-up work. 

We can envisage further GOOFy family symmetries: it is a simple matter to make a version of the Peccei-Quinn $U(1)$ model,
which would look like the $Z_2$ one we just discussed, minus the $m^2_{12}$ and $\lambda_5$ coefficients. That model has some
very interesting aspects, and a version of it will be the subject of a dedicated paper to come out shortly. A GOOFy version of the
$SO(3)$ 2HDM is also possible (with only the couplings $\lambda_1$ and $\lambda_3$ surviving the symmetry imposed), but since no
realistic Yukawa interactions were ever found for that model we will not further consider it.

\section{Conclusions: no longer goofy?}
\label{sec:conc}

In this paper we explained how the imaginary scaling transformations of fields which generate the so-called GOOFy symmetries
in the scalar and gauge sectors may be extended to fermions. We showed how the procedure is consistent, with the field 
transformations needed to maintain invariance of some interaction terms also leaving invariant all other terms in the 
lagrangian, including gauge kinetic ones. We were able in this manner to re-obtain the CP2 and CP3 Yukawa textures 
which had been shown to preserve the GOOFy parameter relations in~\cite{Ferreira:2023dke} by direct beta-function 
inspection up to two loops. We were also able to extend the procedure to other types of 2HDM, obtaining the relations
that the Yukawa matrices must obey for GOOFy invariance. The fermionic field GOOFy transformations were shown to be 
similar to generalized CP or to Higgs-family transformations -- the former transform fields into linear combinations of 
field complex conjugates, the latter into linear combinations of the fields themselves. In both cases, the hallmark of 
a GOOFy transformation is needed: to wit, the fields and their complex conjugates transform ``independently", which is to
say, if a scalar field  has a given symmetry transformation,  $\cal{F} \rightarrow \cal{G}$, 
the transformation of its complex conjugate  will involve an extra minus sign, which would not be present if one were to take
the conjugate of the previous relation, {\em i.e.}  $\cal{F}^* \rightarrow -\,\cal{G}^*$; a similar thing is needed for 
fermionic GOOFy transformations, but instead of a minus sign one has a factor of $i$. Notice that more general 
transformations are proposed in~\cite{Trautner:2025yxz}, which do not leave the gauge-kinetic terms invariant. The field transformations 
considered in this work, along with the imaginary scaling of the spacetime coordinates, ensure the invariance of the full
lagrangian, and indeed of the action as well~\cite{Ferreira:2025ate}.

Other than verifying the consistency of the procedure, and explaining the CP2 and CP3 Yukawa textures identified as
GOOFy-invariant in~\cite{Ferreira:2023dke}, we were able to obtain two versions of the 2HDM, invariant under GOOFy 
symmetries extended to the fermion sector which are phenomenologically viable. These are different from all GOOFy models
studied in both \cite{Ferreira:2023dke} or~\cite{Grzadkowski:2026gkx}, though their scalar sectors had been listed
in~\cite{Trautner:2025yxz}. These models are: (a) a GOOFy CP1 model with explicit CP violation wherein 
the only quadratic parameter in the scalar potential is the imaginary part of $m^2_{12}$ and the Yukawa couplings are
generic real $3\times 3$ matrices; (b) a GOOFy $Z_2$ model, with explicit CP violation originated by the same form of the 
quadratic scalar terms as in the previous model, but with the usual Yukawa matrices as the usual $Z_2$ models, with Type-I,
Type-II, etc., realizations of the symmetry possible and RG-invariant to all orders. In both cases, and unlike the
CP2 and CP3 realizations suggested by the results of~\cite{Ferreira:2023dke}, the models offer realistic fermionic extensions.
Indeed, both models herein proposed have interesting phenomenology which will be probed in the next years at LHC. To begin 
with, in both models the extra scalars are predicted to have masses limited by unitarity bounds to 
be lower than roughly 800 GeV. Both models are also predicted to have explicit CP-violation, which means that all spin-0
states are CP-admixtures -- this has measurable consequences for the SM-like Higgs, for instance, and need to take
into account limits stemming from electric dipole moments. And while the GOOFy CP1 model has FCNC in the scalar-fermion
interactions, the GOOFy $Z_2$ model is flavour-conserving. For both models, we have enough parameters to produce 
viable fermionic sectors, where all quark masses and CKM matrix can be reproduced; viable scalar sectors, with a SM-like
Higgs boson present and extra scalars with couplings and masses such as to have evaded detection thus far -- this is a 
consequence of the fact that in both models there is enough freedom to choose appropriately the parameters of the 
scalar mass diagonalization matrix to ensure that the 125 GeV scalar has couplings which are SM-like (within the
current experimental precision). This automatically suppresses the couplings of additional scalars  to $ZZ$ pairs, for 
instance, allowing us to evade that search channel. Similar freedom to suppress the couplings of extra scalars to fermions
is expected. A dedicated study of both models would be very interesting, using the most up-to-date experimental results
to fully constrain their allowed parameter space -- that study is a paper by itself, and we reserve it for future study.
However, the remarkable conclusion we can reach is that it is possible to extend these GOOFy symmetries to the fermion
sector, obtaining a fully RG-invariant model, with Yukawa matrices no more complex than those of a Type-(I,II,X,Y) 
2HDM and a very restricted set of quadratic scalar parameters. Those restrictions on the $m^2_{ij}$ parameters are
not the result of any type of fine-tuning, but rather RG-invariant from the application of a GOOFy symmetry. 

Another topic worth discussing is the impact of the GOOFy symmetries on the model's parameters: notice that all scalar
GOOFy transformations thus far proposed have something in common -- their impact on the quartic couplings of the
scalar potential are indistinguishable from those of a ``regular" 2HDM symmetry. This is a trivial consequence of the
structure of those quartic couplings, all of them of the form $\Phi_i^\dagger \Phi_j \Phi_k^\dagger \Phi_l$. Recall that
the distinction between GOOFy scalar doublet transformations and ``regular" ones is an extra minus sign affecting the 
$\Phi_i^\dagger$, and since the quartic potential will always involve two of these conjugate fields, that minus sign
is, for this sector, irrelevant. Indeed, all GOOFy symmetric models produced thus far have quartic sectors which may be 
reproduced by the six usual 2HDM symmetries~\cite{Ivanov:2007de}, or combinations thereof: for instance, the original
GOOFy model from~\cite{Ferreira:2023dke} had a quartic sector identical to that of a CP2-symmetric 2HDM; the several models
identified in this paper have quartic sectors reproduced by CP1, $Z_2$ or $U(1)$ symmetries; and so forth. Thus RG invariance
of those dimensionless quartic couplings is automatically ensured. Notice that the very same observation can be made for the 
gauge interactions -- the GOOFy procedure described in detail here leaves gauge-scalar and gauge-fermion interactions
invariant and identical to those of 2HDM models with ``normal" interactions, since no allowed vertices between these particles
are eliminated. 

We now observe that the same holds for the dimensionless Yukawa couplings: the procedure outlined in this paper for fermionic 
GOOFy transformations leads to Yukawa matrices which are the same as those imposed by ``regular" 2HDM symmetries, both
GCP or Higgs family ones. To be more thorough, what we mean is: a given GOOFy symmetry leads to quartic coupling restrictions, 
which are the same ones as those imposed by a ``normal" symmetry $\cal{S}$, and to certain Yukawa matrix textures, which could 
also result from the imposition of the same $\cal{S}$. What distinguishes GOOFy symmetries from usual ones is therefore
their impact on the dimensionful parameters of the scalar potential (or fermionic mass terms, see~\cite{Trautner:2025prm,Trautner:2025yxz,deBoer:2025jhc}) -- as all GOOFy models thus far found, the restrictions on these
$m^2_{ij}$ parameters cannot be reproduced by regular 2HDM symmetries. This automatically explains, and indeed implies, 
the RG invariance discovered for the dimensionless couplings in GOOFy models: the restrictions found for those couplings
are identical to those found for normal symmetries, and the differences found for the dimensionful parameters are tantamount
to a soft breaking which, by definition, will not affect the renormalization behaviour of dimensionless parameters.
\emph{RG-invariance of GOOFy relations between dimensionless gauge, Yukawa and scalar couplings is therefore assured, since
such relations are identical to those obtained by normal symmetries.}

The new, GOOFy-specific, aspect is the RG invariance of the unusual relations between quadratic parameters themselves.
We used a general argument in section~\ref{sec:omod}, analysing the structure of all-orders $\beta$-functions emerging
from both dimensional analysis and basis-transformation properties. But with hindsight, the quadratic coupling 
RG-invariance relations found may even seem obvious. For instance, in the GOOFy CP1 example
treated in this paper, the scalar and Yukawa couplings are all real, the quadratic terms are $m^2_{11} = m^2_{22} = 0$
and $m^2_{12} = \pm\, i\, |m^2_{12}|$; then, dimensional analysis tells us that any loop contribution to  $m^2_{11}$ 
or $m^2_{22}$ would have to be proportional to $m^2_{12}$ multiplied by dimensionless couplings -- and since those couplings
are all real, the resulting contribution would be complex, which is absurd. Thus no $m^2_{11}$, $m^2_{22}$ are generated 
at any loops, and the same argument also shows that the phase of $m^2_{12}$ cannot be altered by radiative contributions. 
A similar reasoning holds for the GOOFy $Z_2$ model studied here: all quartic couplings are real and therefore cannot 
generate, from a purely imaginary $m^2_{12}$, non-zero values for $m^2_{11}$, $m^2_{22}$ at any loop. And since the
(complex) Yukawa matrices are those of Type-I (the same conclusion holds for the other flavour conserving Types) 
fermions interact only with $\Phi_2$ but they \textit{also} cannot generate non-zero values for 
$m^2_{11}$, $m^2_{22}$. The argument goes like this: any diagram contributing to the renormalization of $m^2_{11}$
(an analogous argument holds for $m^2_{22}$)
will have incoming and outgoing $\Phi_1$ legs; it will need to include, in some line of the diagram, a $m^2_{12}$ 
{\em mass insertion}, which converts a $\Phi_1$ line into a $\Phi_2$ one; and since only $\Phi_2$ couples to fermions
it will be impossible to ever turn this into a $\Phi_1$ line to ``close" the diagram. The absence of $\lambda_6$ and
$\lambda_7$ quartic couplings leads to the same conclusion. But even if one finds these RG-invariances obvious 
\textit{a posteriori}, the fact remains that they were only found through GOOFy symmetry implementations.

Which leads us to the final point to stress out: whether these RG-invariant relations are the result of a symmetry
or not, they exist and lead to interesting, testable phenomenology; to possible explanations of the hierarchy
question; to new versions of the 2HDM with parameter choices which are not fine-tuning, but rather renormalization
invariant to all orders. And only through the GOOFy procedure can they be found, taking the transformations of 
fields and their conjugates as independent. Different interpretations of this procedure are available: (A) the original
one from~\cite{Ferreira:2023dke} and this paper considers an imaginary scaling of doublet components, 
gauge and fermion fields and, in order to obtain invariance of kinetic terms, spacetime coordinates are also
taken to scale with a factor of $i$; this process is remarkably consistent, and shown~\cite{Ferreira:2025ate}
to be able to be used for the calculation of the 1-loop effective potential. (B) The approach
championed in~\cite{Trautner:2025prm,deBoer:2025jhc,Trautner:2025yxz} abandons invariance of the gauge-kinetic terms,
arguing that such GOOFy-violating terms amount to soft breakings of the RG-invariant relations found. 
(C) And finally, the approach of~\cite{Haber:2025cbb} sees GOOFy parameter relations emerging as algebraic 
relations between couplings of theories with a larger field content upon which one imposes an ordinary symmetry; this
approach has thus far only been attempted for simple theories with only scalars, but it is expected to be possible to
generalise it for models with fermions, and perhaps gauge fields as well. All of these approaches have, thus far, 
undesirable features: (C) obtains parameter relations in a theory with higher field content and argues that
they imply features of a theory with less fields, not through decoupling of any degrees of freedom but rather from 
algebraic analogies; (B) uses GOOFy transformations to eliminate parameters in the scalar potential but argues that
terms involving kinetic and gauge interactions which break those transformations behave as soft breaking interactions;
and (A), the approach of this paper, uses imaginary scaling of real fields and spacetime coordinates, which is bizarre
to say the least. One could however argue that other symmetries also go beyond usual transformations of 
fields: in conformal symmetry fields are scaled by real numbers, not transformed by unitary matrices, and in
supersymmetric transformations Grassmann numbers are used, so is  imaginary scaling really that shocking? 

However strange the procedure proposed by the (A) method, and whether it
corresponds to some new type of symmetry or instead is simply an effective way to discover RG-symmetric regions 
of parameter space, the point remains that it has proven to be effective in discovering new, interesting models
(a 3HDM implementation is currently being worked on~\cite{SARA:2026new}). As was proven in this work, the resulting models
have fully sensible scalar, gauge and Yukawa interactions, and the Yukawa sector needs not be very peculiar to 
be GOOFy invariant -- GOOFy Type-I, II, etc models with flavor conservation were identified here. As such, though 
these models may well be called GOOFy due to their strangeness and the names of the authors of~\cite{Ferreira:2023dke}, the
amount of work that has been produced about them and the quality thereof shows, in the author's opinion, that they are 
not goofy at all, and should be taken seriously.

\subsection*{Acknowledgments:} 

The author is supported by \textit{Funda\c c\~ao para a Ci\^encia e a Tecnologia} under contract
UIDP/00618/2025, UIDB/00618/2025  and through the PRR (Recovery and Resilience
Plan), within the scope of the investment “RE-C06-i06 - Science Plus Capacity Building”, measure 
``RE-C06-i06.m02 - Reinforcement of financing for International Partnerships in Science,
Technology and Innovation of the PRR”, under the project with the reference 2024.03328.CERN.  
Useful discussions with Andreas Trautner are graciously acknowledged. 

{\footnotesize
\bibliographystyle{utphys}
\bibliography{biblio}
}

\end{document}